\documentclass[11pt,a4paper]{article}
\usepackage[utf8]{inputenc}
\usepackage[english]{babel}
\usepackage{amsmath}
\usepackage{amsfonts}
\usepackage{amssymb}
\usepackage[pdftex]{graphicx}
\usepackage[font=small,labelfont=bf]{caption}
\usepackage{epstopdf}
\usepackage[T1]{fontenc}
\usepackage{etoolbox}
\usepackage{cleveref}
\usepackage{multirow}
\usepackage{amsbsy}
\usepackage[top=1.5in, bottom=1.5in, left=0.8in, right=0.8in]{geometry}
\usepackage{color,soul}
\usepackage{lineno}
\usepackage{tikz}
\usetikzlibrary{patterns}
\usepackage{cleveref}
\usepackage{subcaption}
\usepackage{url}
\usepackage[bottom]{footmisc}
\usepackage{ctable}
\usepackage{enumitem}
\usepackage{wrapfig}
\usepackage{kbordermatrix}
\usepackage{unitsdef}

\allowdisplaybreaks

\makeatletter
\renewcommand\tableofcontents{%
    \@starttoc{toc}%
}
\makeatother

\allowdisplaybreaks

\DeclareUnicodeCharacter{00A0}{~}
\DeclareUnicodeCharacter{2212}{~}

\begin{document}

\newcommand{\ipnp}{\ensuremath{_{i+1}^{n+1}}}
\newcommand{\imnp}{\ensuremath{_{i-1}^{n+1}}}
\newcommand{\inp}{\ensuremath{_{i}^{n+1}}}

\newcommand{\ipn}{\ensuremath{_{i+1}^{n}}}
\newcommand{\imn}{\ensuremath{_{i-1}^{n}}}
\newcommand{\inn}{\ensuremath{_{i}^{n}}}

\newcommand{\np}{\ensuremath{^{n+1}}}
\newcommand{\n}{\ensuremath{^{n}}}

\newcommand{\subij}{\ensuremath{_{i,\,j}}}
\newcommand{\subimj}{\ensuremath{_{i-1,\,j}}}
\newcommand{\subipj}{\ensuremath{_{i+1,\,j}}}
\newcommand{\subijm}{\ensuremath{_{i,\,j-1}}}
\newcommand{\subijp}{\ensuremath{_{i,\,j+1}}}

\newcommand{\erf}{\ensuremath{\text{erf}}}

\newcommand{\etal}{\emph{et al.}$\;$}

\newcommand{\ud}{\mathop{}\!\mathrm{d}}

\newcommand{\red}{\textcolor{red}}
\newcommand{\blue}{\textcolor{blue}}

\newcommand{\highlight}[1]{%
  \colorbox{gray!25}{$\displaystyle#1$}}

\newcommand\T{\rule{0pt}{5ex}}

\providecommand{\keywords}[1]{\hspace{2.5mm} \textbf{Keywords} #1}

\newcommand*\samethanks[1][\value{footnote}]{\footnotemark[#1]}

\makeatletter
\let\@fnsymbol\@arabic
\makeatother
\title{Mathematical model for substitutional binary diffusion in solids}

\renewcommand{\thefootnote}{\arabic{footnote}}

\author{H. Ribera \thanks{Centre de Recerca Matemàtica, Campus de Bellaterra, Edifici C, 08193 Bellaterra, Barcelona, Spain.}$\,\,^{,}\,$\thanks{Department de Matemàtica Aplicada, Universitat Politècnica de Catalunya, Barcelona, Spain.}$ \, ^{,}$ \footnote{Corresponding author} \and B.~R. Wetton \thanks{Mathematics Department, University of British Columbia, Vancouver BC, Canada V6T 1Z2.} \and T.~G. Myers \footnotemark[1]$\,\,^{,}\,$\footnotemark[2]}

\date{\today}

\maketitle


\begin{abstract}

In this paper we detail the mechanisms that drive substitutional binary diffusion and derive appropriate governing equations. We focus on the one-dimensional case with insulated boundary conditions. Asymptotic expansions are used in order to simplify the problem. We are able to obtain approximate analytical solutions in two distinct cases: the two species diffuse at similar rates, and the two species have largely different diffusion rates. A numerical solution for the full problem is also described.

\end{abstract}

\keywords{Kirkendall effect $\cdot$ Mathematical model $\cdot$ Substitutional diffusion}


%
%
%


\section{Introduction}

The first systematic study on solid state diffusion was carried out by Roberts \cite[Part II]{Roberts1896} in 1896, in which he studied diffusion of gold into solid lead at different temperatures, and that of gold into solid silver. 
At the time it was believed that in binary diffusion both species had the same diffusion rate. Pfeil \cite{Pfeil1929} noticed some strange behaviour in iron/steel oxidation in which muffle pieces would fall into the surface of the oxidising iron and were slowly buried until they disappeared beneath the surface. By breaking up the oxidised layer these muffle pieces could be recovered. This seemed to indicate that the diffusion rate of iron and oxygen were not the same. Motivated by this observation Smigelskas and Kirkendall \cite{Smigelskas1947} designed an experiment in which a rectangular block of brass (Cu-Zn alloy) was wound with molybdenum (Mo) wires, since Mo is inert to the system and moves only depending on the transferred material volume. This block was then electroplated with pure Cu, and afterwards the resulting block was annealed at 1058 K. They found that the Mo wires had moved from their original position, which could only mean that Cu and Zn had different diffusion rates. Moreover, this changed the way solid diffusion was understood, since, the now called \emph{Kirkendall effect} showed evidence of a vacancy diffusion mechanism instead of substitutional or ring mechanism, which were the ones believed to be driving binary diffusion in alloys at the time.

Vacancy sites are defects in the lattice, and are basically lattice sites which should be occupied by an atom if the crystal structure was perfect. Atoms use these empty lattice spaces to diffuse. A consequence of the Kirkendall effect is the fact that voids may form and in metals this implies deterioration in their mechanical, thermal and electrical properties. It should be noted that Huntington and Seitz \cite{Huntington1942}, five years before Kirkendall's contribution, argued that indeed, it is the vacancy mechanism that drives diffusion, but because of World War II their work was overlooked. 

Sometimes when there are many vacancies in an alloy a nonequilibrium situation occurs and they supersaturate. They can either form $F$ sinks or $K$ sinks, which means that either a void is created by the union of many vacancies or that the vacancies join but do not form a void, respectively {\cite{Gusak2013}}. Further, grain boundaries and dislocations serve as sources and sinks of vacancies, though in alloys their efficiency as source/sink decreases. Thus allowing this nonequilibrium situations. Recently, this has lead to 
the Kirkendall effect being used to create hollow nanostructures, although the first example of using the Kirkendall effect to create hollow structures was by Aldinger \cite{Aldinger1974}. 
Hollow nanocrystals were first produced by Yin \etal \cite{Yin2004}. 
Gonzalez \etal \cite{Gonzalez2011} were able to synthesise different shapes of nanostructures such as spheres, cubes and tubes at room temperature. 
This type of structure has many possible applications. In biomedicine, they can be used for simultaneous diagnosis and therapy, and the hollow inside can be used to transport drugs and biomolecules and then release them in a controlled manner \cite{An2009}. Piao \etal \cite{Piao2008} used hollow nanocapsules of magnetite that not only were used as a drug delivery vehicle but as a T$_2$ magnetic resonance imaging contrast (MRI) agent. A full review on the nanostructures and MRI can be found in \cite{Na2009}. In the lithium-ion batteries context they have been proposed to be included in the electrodes to enhance rate capability and cycling stability \cite{Wang2012}. Hollow nanoparticles have also been reported to be good catalysts \cite{Kim2002,Li2006}. A review on synthesis and applications of hollow nanostructures can be found in \cite{Lou2008}.

Mathematical models have been proposed to explain the Kirkendall effect. One of the first analysis presented to study this phenomena was the one by Darken {\cite{Darken1948}}, which does not include vacancies. The works of Morral \emph{et al.} {\cite{Morral2001}} and Wierzba {\cite{Wierzba2013}}, which is based in Darken's analysis, also study the Kirkendall effect, but do not mention vacancies in their studies. See {\cite{Gusak2010}} for a discussion of various models. Fan \etal \cite{Fan2007} describe a theory of the physics behind the Kirkendall effect, consisting of two stages: the first one involves the creation of small voids in the compound interface via bulk diffusion, and the second one which strongly relies on surface diffusion of the core material. They say that this model works for both nanospheres and nanotubes. Yu \etal \cite{Yu2007} present a model with vacancy sources and sinks and solve it numerically. Jana \etal \cite{Jana2013} create hollow nanoparticles and present a mathematical model that aims to capture the observed phenomena. The results of the model match the experimental data, but in it there is a free parameter chosen for that purpose. Furthermore, the boundary conditions do not seem to match the physical description of the problem.

In this paper we rigorously derive the governing equations for a substitutional binary diffusion problem and make sensible assumptions to reduce them in order to have an analytically tractable problem. We will pose a 1D problem, the simplest scenario possible in order to gain insight into the physics behind the Kirkendall effect, with the future goal in mind of being able to model the creation of hollow nanostructures.

In the following section we will present expressions for the fluxes of the two species in a binary diffusion problem. In doing so, we obtain expressions for the concentration dependent diffusion coefficients. We will then write the fluxes in terms of the fast diffuser and vacancies, since keeping track of the latter is crucial for our goal. After obtaining the governing equations for the problem we will use them in the development of a one dimensional test case in Section \ref{sec-test-casel}. In two limiting cases we can use asymptotic expansions to simplify the problem and give analytical solutions. These cases correspond to assuming that one species is much faster than the other or that they both diffuse at almost the same rate. We also provide a numerical solution to the full problem. In the results section we demonstrate that the analytical reduction of the diffusion coefficients is valid and thus the reduced governing equations can be used to treat this problem.

\section{Substitutional diffusion}
\label{sec-subs-diffuion}

Consider a binary crystalline solid composed of three species: atomic species A, atomic species B, and vacancies V. We label the fast diffuser as species A, and the slow one B. The driving forces for the diffusion of species are the gradients of the chemical potentials $\mu_i$, so the concentration fluxes of the components of the system are \cite{Manning1961}
\begin{align}
\label{fluxes-1}
    J_A &= -L_{AA} \nabla \mu_A -L_{AB} \nabla \mu_B -L_{AV} \nabla \mu_V, \\
\label{fluxes-2}
    J_B &= -L_{BA} \nabla \mu_A -L_{BB} \nabla \mu_B -L_{BV} \nabla \mu_V, \\
\label{fluxes-3}
    J_V &= -L_{VA} \nabla \mu_A -L_{VB} \nabla \mu_B -L_{VV} \nabla \mu_V,
\end{align}
where $L_{ij}$ are the kinetic transport coefficients. It holds that $L_{AB} = L_{BA}$, $L_{AV} = L_{VA}$, and $L_{BV} = L_{VB}$ \cite{Onsager1931}. In a perfect lattice region (free of dislocations, grain boundaries and surfaces) lattice sites are conserved, so
\begin{equation}
\label{fluxes-perfect-lattice}
J_A + J_B + J_V = 0.
\end{equation}
Substituting for the fluxes from \eqref{fluxes-1}-\eqref{fluxes-3} and equating the coefficients of the different chemical potentials leads to relations $L_{VA} = -(L_{AA}+L_{AB})$, $L_{VB} = -(L_{AB} + L_{BB})$ and $L_{VV} = -(L_{AV}+L_{BV})$. It also means that we only need two of the fluxes to fully define the system,
\begin{align}
	\label{fluxes-A-B-1}
	J_A &= -L_{AA} \nabla (\mu_A - \mu_V) -L_{AB} \nabla (\mu_B - \mu_V), \\
	\label{fluxes-A-B-2}
	J_B &= -L_{AB} \nabla (\mu_A - \mu_V) -L_{BB} \nabla (\mu_B - \mu_V).
\end{align}
We write in terms of the fluxes of A and B to illustrate the fact that substitutional diffusion of an atom in a perfect lattice structure occurs via positional exchange with a neighbouring site.

The kinetic transport coefficients $L_{ij}$ are defined as \cite{Manning1967,Moleko1989}
\begin{align}
	\label{Moleko-KTC-1}
	L_{AA} &= X_V X_A \Gamma_A \frac{\rho \lambda a^2}{k_B T}  \left(1 - \frac{2 X_B \Gamma_A}{\Lambda} \right), \\
	\label{Moleko-KTC-2}
    L_{AB} &= L_{BA} = \Gamma_A \Gamma_B X_A X_B X_V \frac{2\rho \lambda a^2}{k_B T \Lambda}, \\
	\label{Moleko-KTC-3}
    L_{BB} &= X_V X_B \Gamma_B \frac{\rho \lambda a^2}{k_B T}  \left(1 - \frac{2 X_A \Gamma_B}{\Lambda} \right),
\end{align}
where $X_i$ are the mole fractions corresponding to the $i$-th species (and, by the definition of mole fraction, $X_A + X_B + X_V = 1$), $\rho$ is the lattice site density, $\lambda$ is a geometric factor that depends on crystal structure, $a$ is the atomic hop distance, $k_B$ is the Boltzmann constant, $T$ is the temperature of the system, and $\Gamma_A$ and $\Gamma_B$ are the jump frequencies of species A and B, respectively. The jump frequency is the rate at which atoms jump to an adjacent available site. We denote $\Gamma = \Gamma_A/\Gamma_B > 1$, since A is the fast diffuser. A list of typical values of these parameters is given in Table \ref{table-parameter-values}. If species $i$ is absent then the coefficients $L_{ij}$ are such that $J_i = 0$.  The parameter $\Lambda$ is defined as \cite{Moleko1989}
\begin{equation}
\label{Lambda-Moleko}
\begin{split}
\Lambda = &\frac{1}{2}(F_0+2)(X_A\Gamma_A + X_B\Gamma_B) - \Gamma_A - \Gamma_B + 2(X_A\Gamma_B + X_B\Gamma_A)+ \\
& \sqrt{\left(\frac{1}{2}(F_0+2)(X_A\Gamma_A + X_B\Gamma_B) - \Gamma_A - \Gamma_B\right)^2 + 2F_0\Gamma_A\Gamma_B},
\end{split}
\end{equation}
where $F_0 = \frac{2f_0}{1-f_0}$, and $f_0$ is the correlation factor for a single component solid with the crystal structure of the A-B alloy. The fluxes are now well defined.

In order to derive explicit expressions for the fluxes involving the mole fractions of the atomic species, we define the new chemical potentials $\tilde{\mu}_A = \mu_A - \mu_V$ and $\tilde{\mu}_B = \mu_B - \mu_V$ which may be written as \cite{Yu2007} 
\begin{equation}
\label{chemical-potential-1}
\tilde{\mu}_i = \frac{\partial G(X_A,X_B)}{\partial X_i},
\end{equation}
where $G$ is the Gibbs free energy. According to the ideal mixing condition, the free Gibbs energy per lattice site in the A-B alloy with vacancies is
\begin{equation}
\label{free-Gibbs-energy}
G(X_A,X_B) = k_B T \left[X_A \ln(X_A) + X_B \ln(X_B) + X_V \ln(X_V) \right].
\end{equation}
We can rewrite equations \eqref{fluxes-A-B-1} and \eqref{fluxes-A-B-2} as
\begin{align}
	\label{fluxes-A-B-5}
	J_A &= -\rho \left(L_{AA} \frac{1}{\rho} \frac{\partial \tilde{\mu}_A}{\partial X_A} + L_{AB} \frac{1}{\rho} \frac{\partial \tilde{\mu}_B}{\partial X_A} \right)\nabla X_A - \rho \left(L_{AA} \frac{1}{\rho} \frac{\partial \tilde{\mu}_A}{\partial X_B} + L_{AB} \frac{1}{\rho} \frac{\partial \tilde{\mu}_B}{\partial X_B} \right)\nabla X_B, \\
	\label{fluxes-A-B-6}
	J_B &= -\rho \left(L_{BA} \frac{1}{\rho} \frac{\partial \tilde{\mu}_A}{\partial X_A} + L_{BB} \frac{1}{\rho} \frac{\partial \tilde{\mu}_B}{\partial X_A} \right)\nabla X_A - \rho \left(L_{BA} \frac{1}{\rho} \frac{\partial \tilde{\mu}_A}{\partial X_B} + L_{BB} \frac{1}{\rho} \frac{\partial \tilde{\mu}_B}{\partial X_B} \right)\nabla X_B.
\end{align}
The reason for keeping the $\rho$ factor in this form will become apparent later.

If we define the diffusion coefficients as
\begin{equation}
\label{diffusion-coefficients-matrix}
  	\begin{pmatrix}
        D_{AA} & D_{AB} \\
		D_{BA} & D_{BB}
     \end{pmatrix}
   =\begin{pmatrix}
		L_{AA} & L_{AB} \\
		L_{BA} & L_{BB}
  	\end{pmatrix}
    \begin{pmatrix}
		\frac{1}{\rho} \frac{\partial \tilde{\mu}_A}{\partial X_A} & \frac{1}{\rho} \frac{\partial \tilde{\mu}_A}{\partial X_B} \\
		\frac{1}{\rho} \frac{\partial \tilde{\mu}_B}{\partial X_A} & \frac{1}{\rho} \frac{\partial \tilde{\mu}_B}{\partial X_B}
    \end{pmatrix},
\end{equation}
then the fluxes \eqref{fluxes-A-B-5}, \eqref{fluxes-A-B-6} may be written in terms of the mole fractions
\begin{align}
\label{fluxes-A-B-7}
J_A &= -\rho D_{AA} \nabla X_A - \rho D_{AB} \nabla X_B,\\
\label{fluxes-A-B-8}
J_B &= -\rho D_{BA} \nabla X_A - \rho D_{BB} \nabla X_B.
\end{align}
Note that the diffusion coefficients are concentration functions via the chemical potentials and the kinetic transport coefficients.

\subsection{Fluxes in terms of the fast diffuser and vacancies}

In binary diffusion it is the concentration of A and B that are of practical interest. However, the process is only possible due to the presence of vacancies. The vacancy concentration is typically 6 orders of magnitude smaller than $X_A$ or $X_B$, so the details of the evolution of $X_V$ are easily lost in a numerical solution. For this reason, from now on we will work with the fast diffuser A and the vacancies. This means working with the two fluxes $J_A$ and $J_V$. Since lattice sites are conserved, $\sum_i X_i = 1$, we may write
\begin{equation}
\label{molefracs-perfect-lattice}
\nabla X_A + \nabla X_B + \nabla X_V = 0.
\end{equation}
Using equations \eqref{fluxes-perfect-lattice}, \eqref{fluxes-A-B-7}, \eqref{fluxes-A-B-8} and \eqref{molefracs-perfect-lattice}, we then obtain
\begin{align}
	\label{fluxes-B-V-3}
	J_A &= -\rho D_{AA}^V \nabla X_A + \rho D_{AV} \nabla X_V, \\
	\label{fluxes-B-V-4}
	J_V &= \rho D_{VA} \nabla X_A - \rho D_{VV} \nabla X_V,
\end{align}
where the modified diffusion coefficients are
\begin{align}
\begin{split}
\label{D-vacancies-1}
D_{AA}^V &= D_{AA} - D_{AB}, \;\;\;\;\;\;\;\;\;\;\;\;\;\;\;\;\;\;\;\;\;\;\;\;\;\;\;\;\;\;\;\;\;\;\;\;\;\;\;\;\;\;\;\;\; D_{AV} = D_{AB},\\
D_{VA} &= D_{BA} + D_{AA} - D_{BB} - D_{AB}, \;\;\;\;\;\;\;\;\;\;\;\;\;\;\;\;\;\;\;\;\; D_{VV} = D_{BB} + D_{AB}.
\end{split}
\end{align}

Fick's second law states that the rate of change of concentration in time is equal to the divergence of the flux. Noting that the concentration may be written in terms of the mole fraction, $C_i = \rho X_i$, we find diffusion equations for $X_A$ and $X_V$,
\begin{align}
\label{gov-eq-ficks-1}
\frac{\partial X_A}{\partial t} &= - \frac{1}{\rho}\nabla \cdot J_A = \nabla \cdot \left( D_{AA}^V \nabla X_A \right) - \nabla \cdot \left( D_{AV} \nabla X_V \right),\\
\label{gov-eq-ficks-2}
\frac{\partial X_V}{\partial t} &= - \frac{1}{\rho} \nabla \cdot J_V = -\nabla \cdot \left( D_{VA} \nabla X_A \right) + \nabla \cdot \left( D_{VV} \nabla X_V \right).
\end{align}
Note that the $\rho$ term has now disappeared in the governing equations.

\subsection{Diffusion coefficients}

The diffusion coefficients defined by equations \eqref{diffusion-coefficients-matrix}, \eqref{D-vacancies-1} are very complex, making it difficult to identify the dominant mechanisms. Consequently we will now analyse the expressions for $D_{AA}$, $D_{AB}$, $D_{BA}$ and $D_{BB}$. Starting with $D_{AA}$ we note that it consists of two terms,
\begin{align}
\begin{split}
L_{AA} \frac{1}{\rho} \frac{\partial \tilde{\mu}_A}{\partial X_A} &= X_V X_A \Gamma_A \frac{\rho \lambda a^2}{k_B T} \left( 1 - \frac{2X_B \Gamma_A}{\Lambda}\right)\frac{1}{\rho} k_B T \left(\frac{1}{X_A} + \frac{1}{X_V}\right) \\
&= \Gamma_A \lambda a^2 \left( 1 - \frac{2X_B \Gamma_A}{\Lambda}\right) \left(X_V + X_A \right), \\
\end{split}\\
L_{AB} \frac{\partial \tilde{\mu}_B}{\partial X_A} &= X_V X_A X_B \Gamma_A \Gamma_B \frac{2 \rho \lambda a^2}{k_B T \Lambda}\frac{1}{\rho} \frac{k_B T}{X_V} =  X_A X_B \Gamma_A \Gamma_B \frac{2 \lambda a^2}{\Lambda},
\end{align}
which leads to
\begin{align}
\label{DAA}
D_{AA} &= \lambda a^2 \Gamma_A \left[\left(1 - \frac{2 X_B \Gamma_A}{\Lambda} \right)(X_V+X_A) + \frac{2 X_A X_B \Gamma_B}{\Lambda} \right].
\end{align}
By a similar process we obtain
\begin{align}
\label{DAB}
D_{AB} &= \lambda a^2 \Gamma_A \left[X_A \left(1 - \frac{2 X_B \Gamma_A}{\Lambda} \right)+ \frac{2 X_A \Gamma_B}{\Lambda} (X_V+X_B) \right],\\
\label{DBA}
D_{BA} &= \lambda a^2 \Gamma_B \left[\frac{2 X_B \Gamma_A}{\Lambda} (X_V+X_A) + X_B \left(1 - \frac{2 X_A \Gamma_B}{\Lambda} \right) \right],\\
\label{DBB}
D_{BB} &= \lambda a^2 \Gamma_B \left[\frac{2X_A X_B \Gamma_A}{\Lambda} + \left(1 - \frac{2 X_A \Gamma_B}{\Lambda} \right) (X_V+X_B) \right].
\end{align}

Substituting \eqref{DAA}-\eqref{DBB} into \eqref{D-vacancies-1} gives
\begin{align}
\label{DAAV}
D_{AA}^V &= \lambda a^2 \Gamma_A X_V \left[ 1 - \frac{2}{\Lambda}\left( \Gamma_A (1-X_V)-X_A(\Gamma_A-\Gamma_B)\right) \right],\\
\label{DAV}
D_{AV} &= \lambda a^2 \Gamma_A X_A \left[ 1 - \frac{2}{\Lambda}\left( (1-X_A)(\Gamma_A-\Gamma_B) - \Gamma_A X_V \right)\right],\\
\label{DVA}
D_{VA} &= \lambda a^2 X_V \left[ (\Gamma_A-\Gamma_B)\left( 1 - \frac{2}{\Lambda}\left( (1-X_A-X_V)\Gamma_A + \Gamma_B X_A \right) \right) \right],\\
\label{DVV}
D_{VV} &= \lambda a^2 \left[ X_A (\Gamma_A-\Gamma_B) \left( 1 - \frac{2}{\Lambda} \left((1-X_A)(\Gamma_A-\Gamma_B)-X_V\Gamma_A \right)\right) + \Gamma_B\right].
\end{align}

The governing equations \eqref{gov-eq-ficks-1}-\eqref{gov-eq-ficks-2} for the diffusion of species A and the vacancies are now well defined.


\vspace{2mm}

\begin{table}[h]
\centering
\begin{tabular}{clcc}
\hline
parameter & name & value & units \\ \specialrule{1.3pt}{1pt}{1pt}

$\rho$ & lattice site density & $6.021 \times 10^{28}$ & atoms/m$^{3}$ \\ \hline

$\lambda$ & geometric factor & $1/6$ & - \\ \hline

$f_0$ & geometric correlation factor & $0.7815$ & - \\ \hline

$F$ & $2 f_0 / (1 - f_0)$ & $7.1533$ & - \\ \hline



$a$ & lattice constant & $4.05 \times 10^{-10}$ & m \\ \hline

$\Gamma_B$ & hopping frequency of slow diffuser & $10^7$ & Hz \\ \hline

$\Gamma$ & ratio of hop frequencies of slow and fast diffuser & - & - \\ \hline

$\Gamma_A$ & hopping frequency of fast diffuser & $\Gamma \Gamma_B$ & Hz \\ \hline

$X_{V,0}$ & initial vacancy mole fraction & $10^{-6}$ & - \\ \hline
\end{tabular}
\caption{Typical parameter values. They are quite similar as the ones corresponding to aluminium being the slow diffuser. Data taken from \cite{Manning1971,Yu2007}.}
\label{table-parameter-values}
\end{table}


\section{One dimensional case}
\label{sec-test-casel}

\begin{figure}[!htb]
\centering
\begin{tikzpicture}
\draw[fill=black!20!white] (-5,0) -- (0,0) -- (0,3) -- (-5,3) -- (-5,0);
\draw[fill=white] (0,0) -- (5,0) -- (5,3) -- (0,3) -- (0,0);

\node at (-2.5,1.5) {\Huge A};
\node at (2.5,1.5) {\Huge B};

\draw[arrows=->] (4.2,3.2) -- (4.9,3.2);
\node at (5.2,3.2) {$x$};
\node at (5.2,-0.4) {$l$};
\node at (-5.2,-0.4) {$-l$};


\draw[arrows=->] (-1,1.75) -- (1,1.75);
\draw[arrows=<-] (-0.5,1.25) -- (0.5,1.25);
\end{tikzpicture}
\caption{Sketch of the one dimensional bar case.}
\label{figure-model-drawing}
\end{figure}
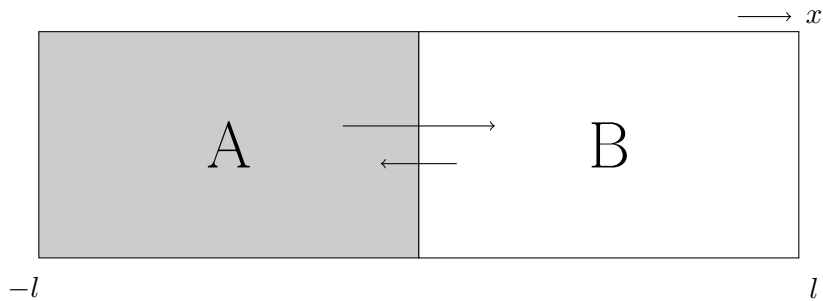

Consider an insulated one-dimensional bar of length $2l$. At $t=0$ the side $x \in [-l,0]$ is made of material A (and a proportion of vacancies), and the side $x \in [0,l]$ is made of material B (and a proportion of vacancies). A sketch of the situation is shown in Figure \ref{figure-model-drawing}.

For $t >0$ the diffusion of species is defined by the 1D forms of \eqref{gov-eq-ficks-1}-\eqref{gov-eq-ficks-2}
\begin{align}
\label{gov-eq-1-mathmodel}
\frac{\partial X_A}{\partial t} &= \frac{\partial}{\partial x}\left(D_{AA}^V \frac{\partial X_A}{\partial x} \right) - \frac{\partial}{\partial x}\left(D_{AV} \frac{\partial X_V}{\partial x} \right),\\
\label{gov-eq-2-mathmodel}
\frac{\partial X_V}{\partial t} &= - \frac{\partial}{\partial x}\left(D_{VA} \frac{\partial X_A}{\partial x} \right) + \frac{\partial}{\partial x}\left(D_{VV} \frac{\partial X_V}{\partial x} \right),
\end{align}
with the diffusion coefficients given in \eqref{DAAV}-\eqref{DVV}, subject to boundary conditions
\begin{equation}
\label{bc}
\frac{\partial X_A}{\partial x}\bigg|_{x=\pm l} = \frac{\partial X_V}{\partial x}\bigg|_{x=\pm l} = 0.
\end{equation}
The boundary conditions confine the material to $x \in [-l,\,l]$. In practice the Kirkendall effect can cause the boundaries to move. We will not study this situation here. The initial conditions are
\begin{equation}
\begin{aligned}
\label{ic}
X_A(x,0) = \left\{
     \begin{array}{lcr}
      X_{A,\text{ini}} & \text{if} & -l < \,\, x < 0,\\
      0 & \text{if} & 0 < \,\, x < l,
     \end{array}
   \right. \;\;\;&\;\;\;
X_B(x,0) = \left\{
     \begin{array}{lcr}
      0 & \text{if} & -l < \,\, x < 0,\\
      X_{B,\text{ini}} & \text{if} & 0 < \,\, x < l,
     \end{array}
   \right. \\
   \; \\
   X_V(x,0) &= X_{V,\text{ini}},
\end{aligned}
\end{equation}
where $X_{A,\text{ini}}$, $X_{B,\text{ini}}$, and $X_{V,\text{ini}}$ denote the constant initial mole fractions of material A, B, and vacancies, respectively, and $X_{i,\text{ini}} = 1 - X_{V,\text{ini}}$, $i=$ A, B.

We now non-dimensionalise the variables
\begin{equation}
\label{nondimensional}
\hat{x} = \frac{x}{l}, \qquad \hat{t} = \frac{\bar{D}_{BB}}{l^2}t,
\end{equation}
where $\bar{D}_{BB} = \lambda a^2 \Gamma_B$. We also rescale $\hat{\Gamma}_i = \Gamma_i/\Gamma_B$. Immediately dropping the hats the governing equations become
\begin{align}
\label{gov-eq-1-mathmodel-ND-full}
\frac{\partial X_A}{\partial t} &= \frac{\partial}{\partial x}\left(D_{AA}^V \frac{\partial X_A}{\partial x} \right) - \frac{\partial}{\partial x}\left(D_{AV} \frac{\partial X_V}{\partial x} \right),\\
\label{gov-eq-2-mathmodel-ND-full}
\frac{\partial X_V}{\partial t} &= - \frac{\partial}{\partial x}\left(D_{VA} \frac{\partial X_A}{\partial x} \right) + \frac{\partial}{\partial x}\left(D_{VV} \frac{\partial X_V}{\partial x} \right),
\end{align}
where
\begin{align}
\label{DAAV-ND}
D_{AA}^V &= \Gamma X_V \left[ 1 - \frac{2}{\Lambda}\left( \Gamma (1-X_V)-X_A(\Gamma-1)\right) \right],\\
\label{DAV-ND}
D_{AV} &= \Gamma X_A \left[ 1 - \frac{2}{\Lambda}\left( (1-X_A)(\Gamma-1) - \Gamma X_V \right)\right],\\
\label{DVA-ND}
D_{VA} &= X_V \left[ (\Gamma-1)\left( 1 - \frac{2}{\Lambda}\left( \Gamma(1-X_V) - X_A(\Gamma-1) \right) \right) \right],\\
\label{DVV-ND}
D_{VV} &=  X_A (\Gamma-1) \left[ 1 - \frac{2}{\Lambda} \left((1-X_A)(\Gamma-1)-X_V\Gamma \right)\right] +1.
\end{align}

The boundary conditions are unchanged (although now applied at $x = \pm 1$).

\subsection{Approximate solutions}

We wish to solve the problem defined by \eqref{gov-eq-1-mathmodel-ND-full}-\eqref{gov-eq-2-mathmodel-ND-full} and appropriate boundary conditions. In order to do so we try to simplify the problem. A sensible assumption to make is that $X_V = \mathcal{O}(\epsilon)$, where $\epsilon \ll 1$ is taken to be the initial vacancy mole fraction. We write $X_A = X_{A,0} + \epsilon X_{A,1}$ and $X_V = \epsilon X_{V,1}$. The governing equations \eqref{gov-eq-1-mathmodel-ND-full}-\eqref{gov-eq-2-mathmodel-ND-full} to first order in $\epsilon$  in a regular asymptotic expansion are
\begin{align}
\label{gov-eq-1-ND}
\frac{\partial X_{A,0}}{\partial t} + \epsilon \frac{\partial X_{A,1}}{\partial t} &= \epsilon \frac{\partial}{\partial x}\left(D_{AA,1}^V \frac{\partial X_{A,0}}{\partial x} \right) - \epsilon \frac{\partial}{\partial x}\left(D_{AV,0} \frac{\partial X_{V,1}}{\partial x} \right),\\
\label{gov-eq-2-ND}
\epsilon\frac{\partial X_{V,1}}{\partial t} &= -\epsilon\frac{\partial}{\partial x}\left(D_{VA,1} \frac{\partial X_{A,0}}{\partial x} \right) + \epsilon\frac{\partial}{\partial x}\left(D_{VV,0} \frac{\partial X_{V,1}}{\partial x} \right).
\end{align}
The four diffusion coefficients come from the expansions
\begin{align}
\begin{split}
\label{DAAV-expand}
D_{AA}^V &= \epsilon D_{AA,1}^V + \mathcal{O}(\epsilon^2), \;\;\;\;\;\;\;\;\;\;\;\;\;  D_{AV} = D_{AV,0} + \epsilon D_{AV,1} + \mathcal{O}(\epsilon^2),\\
D_{VA} &= \epsilon D_{VA,1} + \mathcal{O}(\epsilon^2), \;\;\;\;\;\;\;\;\;\;\;\;\; D_{VV} = D_{VV,0} + \epsilon D_{VV,1} + \mathcal{O}(\epsilon^2),
\end{split}
\end{align}
where
\begin{align}
\label{DAAV-retained-full}
D_{AA,1}^V &= \Gamma X_{V,1} \left[ 1 - \frac{2}{\Lambda}\left( \Gamma(1-X_{A,0}) + X_{A,0} \right) \right],\\
\label{DAV-retained-full}
D_{AV,0} &= \Gamma X_{A,0} \left[ 1 - \frac{2}{\Lambda} (1-X_{A,0})(\Gamma-1) \right],\\
\label{DVA-retained-full}
D_{VA,1} &= X_{V,1} (\Gamma-1) \left[ 1 - \frac{2}{\Lambda}\left( \Gamma (1-X_{A,0}) + X_{A,0} \right) \right],\\
\label{DVV-retained-full}
D_{VV,0} &=  X_{A,0} (\Gamma-1) \left[ 1 - \frac{2}{\Lambda} (1-X_{A,0})(\Gamma-1) \right] +1.
\end{align}

The expression used for  $\Lambda$ in the equations above \eqref{DAAV-retained-full}-\eqref{DVV-retained-full}, to leading order, is
\begin{equation}
\begin{split}
\Lambda = &\frac{1}{2}(F_0+2)(X_{A,0}\Gamma + 1 - X_{A,0}) - \Gamma - 1 + 2\left(X_{A,0} + \Gamma(1-X_{A,0})\right) + \\
&\sqrt{\left( \frac{1}{2}(F_0+2)(X_{A,0}\Gamma + 1 - X_{A,0}) - \Gamma - 1 \right)^2 + 2F_0 \Gamma}
\end{split}
\end{equation}

To first order in $\epsilon$ the problem to solve is now
\begin{align}
\label{new-sys-1-full}
\frac{\partial X_{A,0}}{\partial t} &= 0,\\
\label{new-sys-2-full}
\frac{\partial X_{A,1}}{\partial t} &= \frac{\partial}{\partial x}\left(D_{AA,1}^V \frac{\partial X_{A,0}}{\partial x} \right) - \frac{\partial}{\partial x}\left(D_{AV,0} \frac{\partial X_{V,1}}{\partial x} \right),\\
\label{new-sys-3-full}
\frac{\partial X_{V,1}}{\partial t} &= -\frac{\partial}{\partial x}\left(D_{VA,1} \frac{\partial X_{A,0}}{\partial x} \right) + \frac{\partial}{\partial x}\left(D_{VV,0} \frac{\partial X_{V,1}}{\partial x} \right),
\end{align}
with boundary conditions
\begin{equation}
\label{new-sys-bc-full}
\frac{\partial X_{A,0}}{\partial x}\bigg|_{x=\pm 1} = \frac{\partial X_{A,1}}{\partial x}\bigg|_{x=\pm 1} = \frac{\partial X_{V,1}}{\partial x}\bigg|_{x=\pm 1} = 0.
\end{equation}
Equation \eqref{new-sys-1-full} tells us that in this time scale, $X_{A,0}$ is in steady state and so is defined by the initial condition, $X_{A,0} = f(x) = X_A(x,0)$. 
This indicates the need to study the slow time dynamics;
since on the normal time scale we find that $X_{A,0}$ is constant in time we will not see any behaviour of interest. For this reason, we rescale time as $\tau = \epsilon t$. The problem then becomes, to first order in $\epsilon$,
\begin{align}
\label{gov-eq-1-mathmodel-SlowT}
\frac{\partial X_{A,0}}{\partial \tau} &= \frac{\partial}{\partial x}\left(D_{AA,1}^V \frac{\partial X_{A,0}}{\partial x} \right) - \frac{\partial}{\partial x}\left(D_{AV,0} \frac{\partial X_{V,1}}{\partial x} \right),\\
\label{gov-eq-2-mathmodel-SlowT}
0 &= - \frac{\partial}{\partial x}\left(D_{VA,1} \frac{\partial X_{A,0}}{\partial x} \right) + \frac{\partial}{\partial x}\left(D_{VV,0} \frac{\partial X_{V,1}}{\partial x} \right),
\end{align}
subject to boundary conditions \eqref{new-sys-bc-full}. 

\subsubsection{Special cases}
\label{sec-specialcases}
In the previous section although we could simplify the governing equations we were not able to provide any analytical solutions and thus not much insight into what drives this process. In the following we study two particular cases of the problem in which we are able to find analytical solutions.

\subsubsection*{Case $\Gamma \gg 1$}
\label{sec-GammaB1}

Assuming A diffuses much faster than B, that is, $\Gamma \gg 1$, \eqref{DAAV-retained-full}-\eqref{DVV-retained-full} reduce to
\begin{equation}
\label{reduced-diffusion}
\begin{split}
\;\;\;\;\;\;\;\: D_{AA}^V &\sim \Gamma X_V, \;\;\;\;\;\;\;\;\;\;\;\;\;\;\;\;\;\;\;\;\;\;\;\:\! D_{AV} \sim \Gamma X_A,\\
D_{VA} &\sim \left( \Gamma - 1 \right) X_V, \;\;\;\;\;\;\;\;\;\;\;\;\;\, D_{VV} \sim \left(\Gamma - 1 \right)X_A + 1.
\end{split}
\end{equation}
These reductions hold provided $X_{A,0}$ is not close to zero. The problem becomes
\begin{align}
\label{new-sys-1}
\frac{\partial X_{A,0}}{\partial t} &= 0,\\
\label{new-sys-2}
\frac{\partial X_{A,1}}{\partial t} &= \Gamma X_{V,1} \frac{\partial^2 X_{A,0}}{\partial x^2} - \Gamma X_{A,0} \frac{\partial^2 X_{V,1}}{\partial x^2}, \\
\label{new-sys-3}
\frac{\partial X_{V,1}}{\partial t} &= - (\Gamma-1) X_{V,1} \frac{\partial^2 X_{A,0}}{\partial x^2} + [1+(\Gamma-1)X_{A,0}]  \frac{\partial^2 X_{V,1}}{\partial x^2},
\end{align}
with boundary conditions \eqref{new-sys-bc-full}. The system above leads again to $X_{A,0} = X_A(x,0)$, which is the Heaviside function of $-x$.
Although equation \eqref{gov-eq-2-mathmodel-SlowT} can be integrated it does not give a simple relation between $X_{A,0}$ and $X_{V,1}$ due to the nonlinear diffusion coefficients. Again analytical progress is difficult so we now focus on the slow time dynamics.

\subsubsection*{Slow time dynamics}

Consider the system defined in \eqref{new-sys-1}-\eqref{new-sys-3} and rescale time so that $\tau = \epsilon t$. This leads to
\begin{align}
\label{new-sys-1-tau}
\frac{\partial X_{A,0}}{\partial \tau} &= \frac{\partial}{\partial x} \left( \Gamma X_{V,1} \frac{\partial X_{A,0}}{\partial x}\right) - \frac{\partial}{\partial x} \left( \Gamma X_{A,0} \frac{\partial X_{V,1}}{\partial x} \right), \\
\label{new-sys-2-tau}
0 &= - \frac{\partial}{\partial x} \left( \left(\Gamma - 1\right) X_{V,1} \frac{\partial X_{A,0}}{\partial x}\right) +  \frac{\partial}{\partial x} \left( \left[1 + (\Gamma - 1) X_{A,0} \right] \frac{\partial X_{V,1}}{\partial x} \right).
\end{align}
Integrating \eqref{new-sys-2-tau} yields
\begin{align}
\label{new-sys-3-int-1}
-\left(\Gamma - 1\right) X_{V,1} \frac{\partial X_{A,0}}{\partial x} + \left[1 + (\Gamma - 1) X_{A,0} \right] \frac{\partial X_{V,1}}{\partial x} = 0,
\end{align}
where the constant of integration is zero because of the boundary conditions \eqref{new-sys-bc-full}. Rearranging and integrating by substitution yields
\begin{align}
\label{new-sys-3-int-5}
X_{V,1}(x) &= X_{V,1}(-1) \left( \frac{1 + (\Gamma - 1) X_{A,0}(x)}{1 + (\Gamma - 1) X_{A,0}(-1)} \right),
\end{align}
where $X_{V,1}(-1)$ is picked such that $\int_{-1}^1 X_{V,1}(x)\ud x= 1$. 
Let $M_0 = \int_{-1}^1 X_{A,0}(x)\ud x$. Then equation \eqref{new-sys-3-int-1} can be written as
\begin{align}
\label{new-sys-3-int-8}
X_{V,1}(x) &= \frac{1 + (\Gamma - 1)X_{A,0}(x)}{2 + (\Gamma-1)M_0}.
\end{align}
Substituting \eqref{new-sys-3-int-8} into equation \eqref{new-sys-1-tau} gives
\begin{align}
\label{new-sys-1-tau-2}
\frac{\partial X_{A,0}}{\partial \tau} &=  \frac{\Gamma}{2 + (\Gamma-1)M_0} \frac{\partial^2 X_{A,0}}{\partial x^2}.
\end{align}
This indicates that on a slow time scale the vacancies will adapt to A as described in equation \eqref{new-sys-3-int-8} and A will follow a simple diffusion process described by \eqref{new-sys-1-tau-2}.

Let us define $\alpha = \Gamma/(2 + (\Gamma-1)M_0)$. Solving equation \eqref{new-sys-1-tau-2} subject to \eqref{new-sys-bc-full} is a simple case of separation of variables,
\begin{align}
\label{new-sys-1-tau-2-u-analytical}
X_{A,0}(x,\tau) = \frac{M_0}{2} + \sum_{k=1}^\infty C_k \cos\left( \frac{k \pi}{2} (x+1) \right) e^{ - \left(\frac{k \pi}{2}\right)^2 \alpha \tau },
\end{align}
where
\begin{align}
C_k = \int_{-1}^1 X_{A,0}(x,0) \cos\left( \frac{k \pi}{2} (x+1) \right) \ud x.
\end{align}
The $C_k$ values can be computed analytically for given initial data such as \eqref{ic}. In this case $C_k$ are only nonzero when $k$ is odd,
\begin{equation}
C_k = (-1)^{k+1} \frac{2 X_{A,\text{ini}}}{(2k-1)\pi}.
\end{equation}
Consequently, via equation \eqref{new-sys-3-int-8}, we obtain
\begin{align}
\label{new-sys-1-tau-2-v-analytical}
X_{V,1}(x,\tau) = \frac{\alpha}{\Gamma} \left(1 + (\Gamma - 1) \left[ \frac{M_0}{2} + \sum_{k=1}^\infty C_k \cos\left( \frac{(2k-1) \pi}{2} (x+1) \right) e^{ - \left(\frac{(2k-1) \pi}{2}\right)^2 \alpha \tau } \right]\right).
\end{align}

\subsubsection*{Case $\Gamma \sim 1$}
\label{sec-Gamma1}

Another limit where progress can be made is $\Gamma \sim 1$. This means that both species diffuse at similar rates (although A is still faster). This reduces the diffusion coefficients \eqref{DAAV-retained-full}-\eqref{DVV-retained-full} to
\begin{equation}
\label{reduced-diffusion-G1}
\begin{split}
\;\;\;\;\;\;\;\: D_{AA}^V &\sim \Gamma D_C X_V , \;\;\;\;\;\;\;\;\;\;\;\;\;\;\;\;\;\;\;\;\;\;\;\:\! D_{AV} \sim \Gamma X_A,\\
D_{VA} &\sim \left( \Gamma - 1 \right) D_C X_V , \;\;\;\;\;\;\;\;\;\; D_{VV} \sim \left(\Gamma - 1 \right)X_A + 1,
\end{split}
\end{equation}
where $D_C = 1 - 2/(F0+2)$ is constant. These reductions are valid for all $X_{A}$. We can no longer solve the problem analytically, but these expressions for the diffusion coefficients are much simpler than the original ones, and using them can further simplify the study of the Kirkendall effect.

However, it is possible to make analytical progress if we introduce a small error into the diffusion coefficients $D_{AV}$ and $D_{VV}$ so that
\begin{equation}
\label{reduced-diffusion-G12}
D_{AV} \sim \Gamma D_C X_A, \qquad D_{VV} \sim \left(\Gamma - 1 \right) D_C X_A  + 1,
\end{equation}
where using the parameter values of Table \ref{table-parameter-values}, $D_C = 0.7815$.
As we will see later the errors resulting from this approximation are small. The concentration $X_{A,0}$ is now the same solution as in equation \eqref{new-sys-1-tau-2-u-analytical} but with $\alpha = D_C \Gamma/(2 + (\Gamma-1) D_C M_0)$,
\begin{align}
\label{new-sys-1-tau-2-u-analytical-G1}
X_{A,0}(x,\tau) =  \frac{M_0}{2} + \sum_{k=1}^\infty C_k \cos\left( \frac{(2k-1) \pi}{2} (x+1) \right) e^{ - \left(\frac{(2k-1) \pi}{2}\right)^2 \alpha \tau }.
\end{align}
For the concentration of vacancies $X_{V,1}$ we find
\begin{align}
\label{new-sys-1-tau-2-v-analytical-G1}
X_{V,1}(x,\tau) = \frac{\alpha}{\Gamma D_C} \left(1 + (\Gamma - 1) D_C \left[ \frac{M_0}{2} + \sum_{k=1}^\infty C_k \cos\left( \frac{k \pi}{2} (x+1) \right) e^{ - \left(\frac{k \pi}{2}\right)^2 \alpha \tau } \right]\right).
\end{align}

\subsection{Numerical solution of the slow time dynamics}
\label{sec-numerics}

Let $u_i$ and $v_i$ be the average values of mole fractions $X_{A,0}$ and $X_{V,1}$, respectively, over the interval $(x_{i-1/2},\,x_{i+1/2})$ of length $h$, centered at $x_i$. We introduce the vectors $\mathbf{Y}$ and $\mathbf{Z}$ which represent the interpolated values of $u$ and $v$ falling on the subinterval end points. We have that
\begin{align}
Y_i &= u_{i+1/2} = \frac{u_{i} + u_{i+1}}{2}, & i &= 1,\dots,\,I+1,\\
Z_i &= v_{i+1/2} = \frac{v_{i} + v_{i+1}}{2}, & i &= 1,\dots,\,I+1.
\end{align}

%
%
%
%
%
%

The right-hand side of equation \eqref{gov-eq-1-mathmodel-SlowT} can be discretised as
\begin{equation}
f_i = \frac{1}{h} \left[ (Q_{i} - q_{i}) - (Q_{i-1} - q_{i-1}) \right],
\end{equation}
where
\begin{align}
Q_i &= \Gamma Z_i \left[ 1 - \frac{2}{\Lambda(Y_i)} \left( \Gamma(1-Y_i) + Y_i\right) \right] \frac{u_{i+1}-u_i}{h}, & i &= 1,\dots,\,I+1,\\
q_i &= \Gamma Y_i \left[ 1 - \frac{2}{\Lambda(Y_i)} ( 1-Y_i) (\Gamma - 1) \right] \frac{v_{i+1}-v_i}{h}, & i &= 1,\dots,\,I+1.
\end{align}

The function $\Lambda({Y_i})$ is approximated analytically from \eqref{Lambda-Moleko} as
\begin{equation}
\begin{split}
\Lambda({Y_i}) = &\frac{1}{2}(F_0+2)(Y_i \Gamma + (1-Y_i)) - \Gamma_A - 1 + 2(Y_i + (1-Y_i)\Gamma) \\
& + \sqrt{\left(\frac{1}{2}(F_0+2)(Y_i \Gamma + (1-Y_i)) - \Gamma - 1\right)^2 + 2F_0\Gamma} \;\; + \;\; \mathcal{O}(\epsilon).
\end{split}
\end{equation}

The discretisation of the right-hand side of equation \eqref{gov-eq-2-mathmodel-SlowT} is approximated similarly as
\begin{equation}
g_i = \frac{1}{h} \left[ (-P_{i} + p_{i}) - (-P_{i-1} + p_{i-1}) \right],
\end{equation}
where
\begin{align}
P_i &= Z_i (\Gamma - 1) \left[ 1 - \frac{2}{\Lambda(Y_i)} \left( \Gamma_A(1-Y_i) + \Gamma_B Y_i\right) \right] \frac{u_{i+1}-u_i}{h}, & i &= 1,\dots,\,I+1,\\
p_i &= \left( Y_i (\Gamma - 1) \left[ 1 - \frac{2}{\Lambda(Y_i)} ( 1-Y_i) (\Gamma_A - \Gamma_B) \right] + 1 \right) \frac{v_{i+1}-v_i}{h}, & i &= 1,\dots,\,I+1.
\end{align}

The boundary conditions \eqref{new-sys-bc-full} transform into
\begin{align}
&f_1 = \frac{u_2 - u_1}{h} = 0, \\
&f_{I+2} = \frac{u_{I+2} - u_{I+1}}{h} = 0, \\
&g_1 = \frac{v_2 - v_1}{h} = 0, \\
&g_{I+2} = \frac{v_{I+2} - v_{I+1}}{h} = 0,
\end{align}
where ghost cells $1$ and $I+2$ are introduced outside the domain.

We define the mass matrix $\mathbf{M}$ as
\renewcommand{\kbldelim}{(}
\renewcommand{\kbrdelim}{)}
\[
\mathbf{M} = \kbordermatrix{
  &   &   &  &   &   &   &   \\
1 & 1 & 0 & \cdots & 0 & 0 & \dots & 0 \\
2 & 0 & 1 & \cdots & 0 & 0 & \dots & 0 \\
  & \vdots & \vdots & \ddots & \vdots & \vdots & \ddots & \vdots \\
I & 0 & 0 & \cdots & 1 & 0 & \cdots & 0 \\
I+1 & 0 & 0 & \cdots & 0 & 0 & \dots & 0 \\
I+2 & 0 & 0 & \cdots & 0 & 0 & \dots & 0 \\
\vdots & \vdots & \ddots & \vdots & \vdots & \ddots & \vdots \\
2(I+2) & 0 & 0 & \cdots & 0 & 0 & \cdots & 0
  },
\]
$\mathbf{F} = \left(f_2,\dots,\,f_{I+1}, f_1,\, f_{I+2},\, g_1,\dots,\,g_{I+2} \right)$, and $\mathbf{U} = (\mathbf{u},\mathbf{v})$. Then we can write the problem as
\begin{equation}
\label{DAE-formulation}
\mathbf{M} \frac{\partial \mathbf{U}}{\partial t} = \mathbf{F}(\mathbf{U}).
\end{equation}
The unknowns are $u_i$, $v_i$, for $i = 1,\dots,\, I+2$. Equation \eqref{DAE-formulation} can be solved easily via the ODE routines in MATLAB. We use the \texttt{ode15s} routine.

\section{Results}

In this section we present the results of the one-dimensional case. The parameter values used can be found in Table \ref{table-parameter-values}.

\begin{figure}[!htb]
\makebox[\linewidth][c]{%
\begin{subfigure}[b]{.5\textwidth}
\centering
\includegraphics[width=1\textwidth]{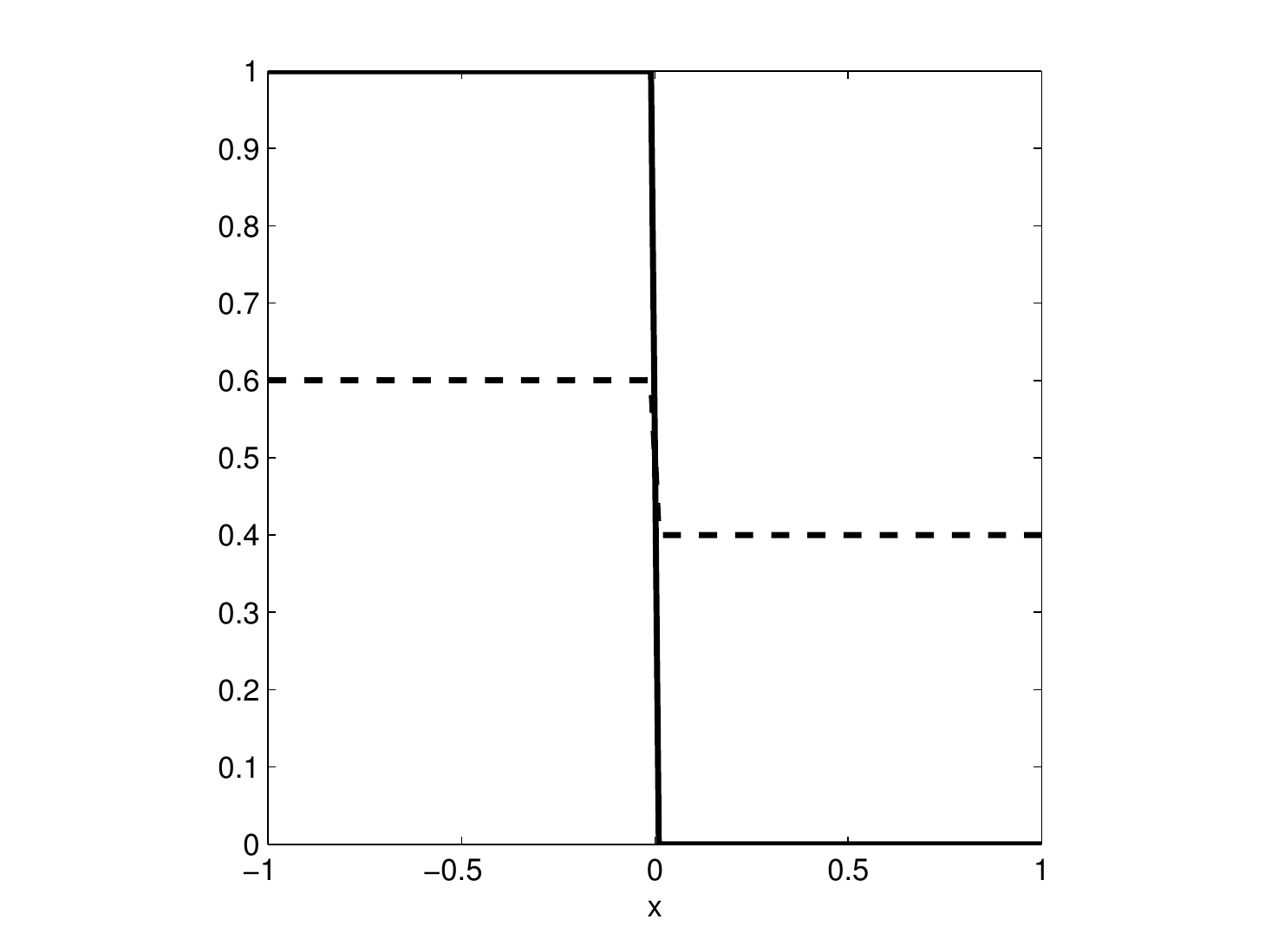}
\caption{$\Gamma = 1.5$.}
\end{subfigure}%
\begin{subfigure}[b]{.5\textwidth}
\centering
\includegraphics[width=1\textwidth]{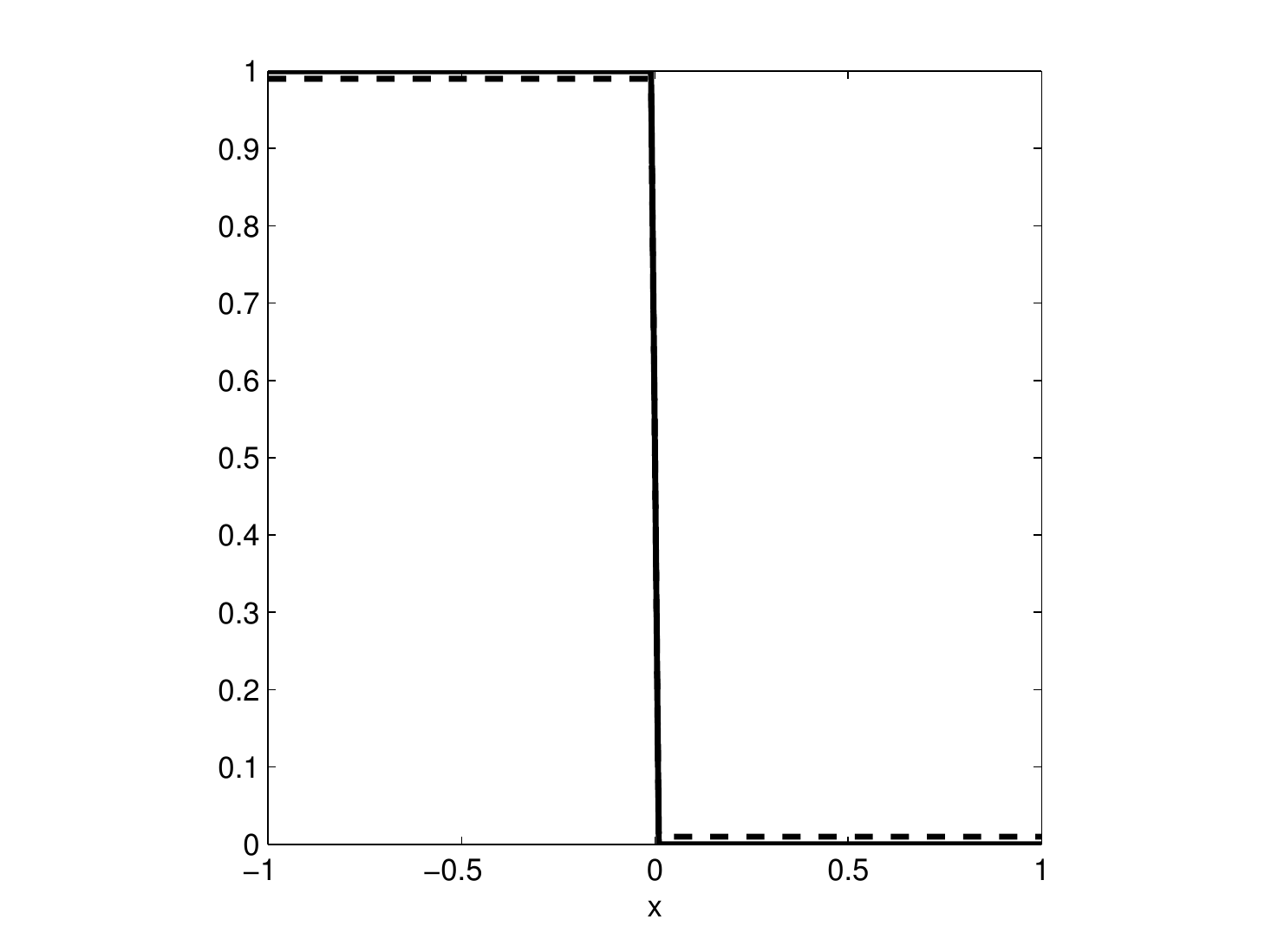}
\caption{$\Gamma = 100$.}
\end{subfigure}%
}
\caption{Solution of equations \eqref{gov-eq-1-mathmodel-SlowT}-\eqref{gov-eq-2-mathmodel-SlowT}. $X_{A,0}$ (solid) and $X_{V,1}$ (dashed) when $\tau \rightarrow 0$ for different $\Gamma$ values.}
\label{figure-FastTime-comparison}
\end{figure}

The fast time system is defined by equations \eqref{new-sys-1-full}-\eqref{new-sys-3-full}. The initial conditions are $X_A(x,0) = H(-x)$ and $X_V(x,0) = 0.5$.
The second time regime corresponds to $t \gg 1$, $\tau = \epsilon t$, and is described by equations \eqref{gov-eq-1-mathmodel-SlowT}-\eqref{gov-eq-2-mathmodel-SlowT}. 
Figure \ref{figure-FastTime-comparison} displays the steady-state (large time) solutions for \eqref{new-sys-1-full}-\eqref{new-sys-3-full}, the two curves represent $X_{A,0}$ (solid) and $X_{V,1}$ (dashed). These provide the initial conditions, $\tau = 0$, for \eqref{gov-eq-1-mathmodel-SlowT}-\eqref{gov-eq-2-mathmodel-SlowT}. Both solutions show that distributions reflect the Heaviside initial condition for $X_A$. On this time-scale there is no noticeable movement of $X_A$, but there has been a significant shift in the vacancies (which indicates that there has been movement of $X_A,\,X_B$ but, since they have a much larger volume fraction it cannot be observed in the figure). The amount of vacancies that move depends strongly on $\Gamma$. When $\Gamma=1.5$, that is A diffuses only slightly faster than B the redistribution of vacancies is relatively small, from the initial value of 0.5 to 0.6 on the left hand side which is balanced by 0.4 on the right hand side. When A diffuses much faster than B, in this example $\Gamma=100$, nearly all vacancies move to the left hand side.
This is easily explained: A and B having a similar jump frequency means that more or less the number of exchanges between A and a vacancy, and B and a vacancy is the same, thus the increase of V on the left hand side of the bar is small.
On the other hand, $\Gamma \gg 1$ means that a large number of A atoms are going to exchange position with vacancies for every B atom that is able to do this type of exchange. The only way to compensate this difference is via vacancy lattice spaces, that end up where A was at the beginning of the process.

\begin{figure}[!htb]
\makebox[\linewidth][c]{%
\begin{subfigure}[b]{.5\textwidth}
\centering
\includegraphics[width=0.94\textwidth]{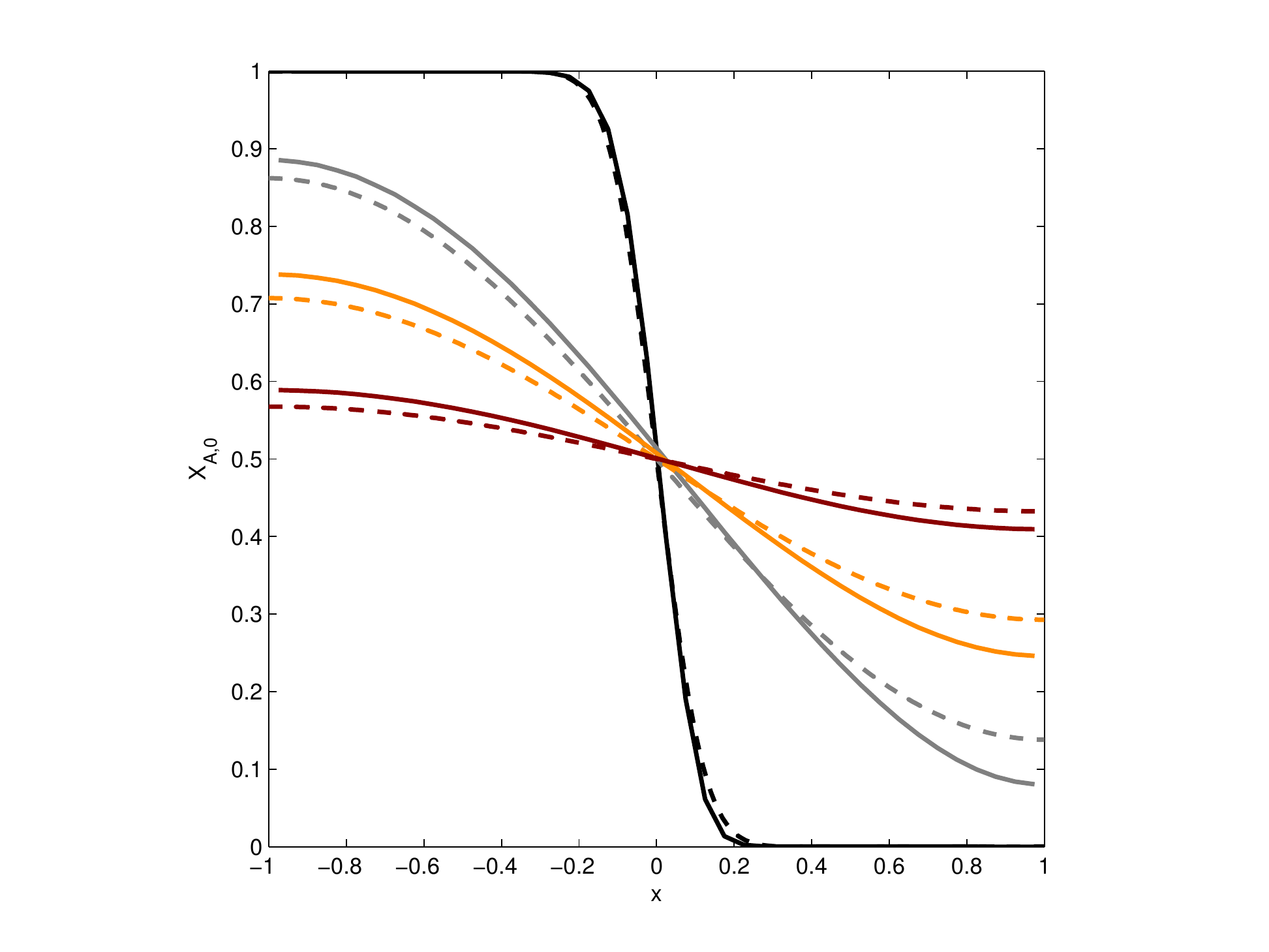}
\caption{$X_{A,0}$ for $\Gamma = 10$.}
\end{subfigure}%
\begin{subfigure}[b]{.5\textwidth}
\centering
\includegraphics[width=0.94\textwidth]{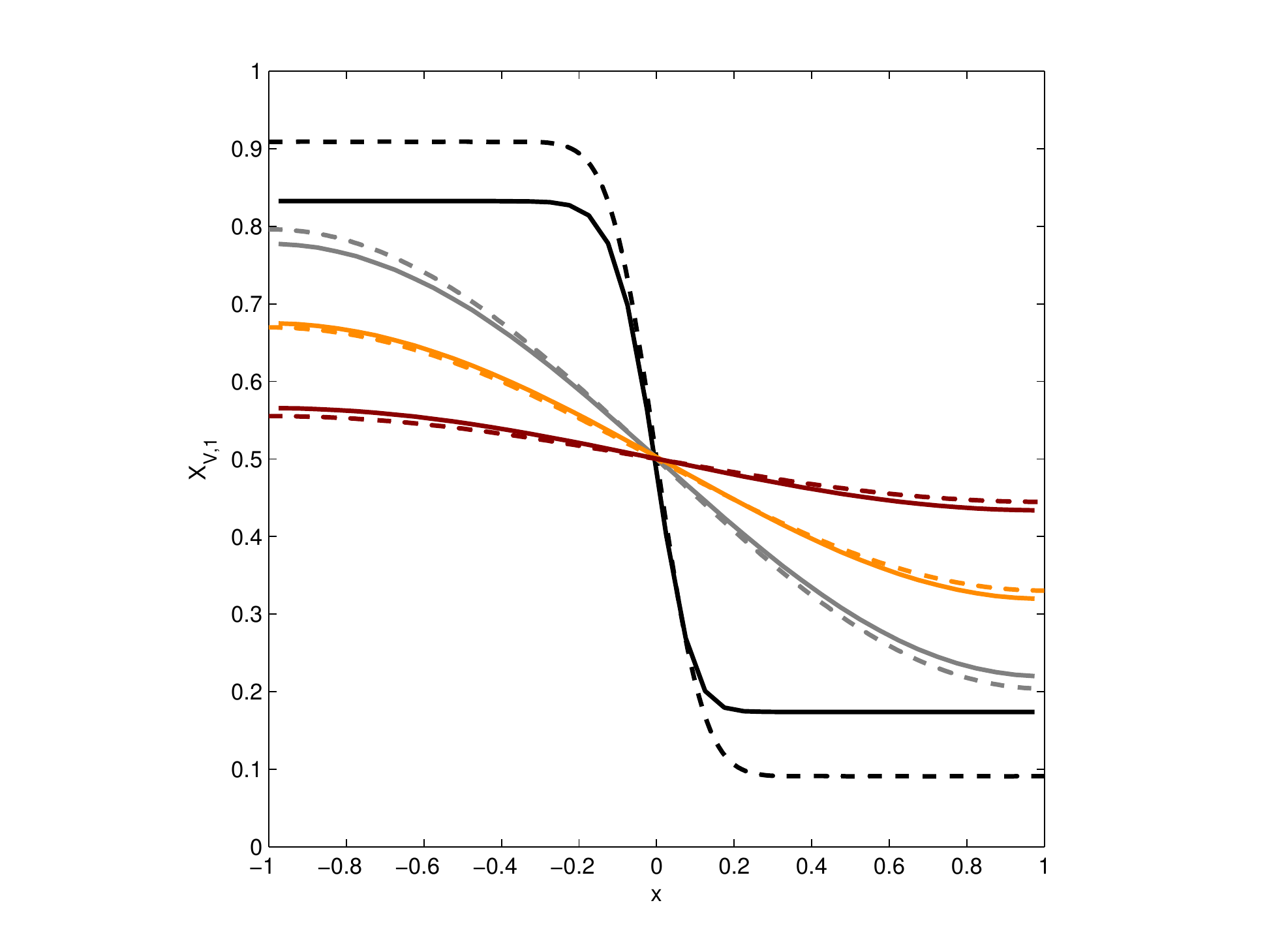}
\caption{$X_{V,1}$ for $\Gamma = 10$.}
\end{subfigure}%
}
\makebox[\linewidth][c]{%
\begin{subfigure}[b]{.5\textwidth}
\centering
\includegraphics[width=0.94\textwidth]{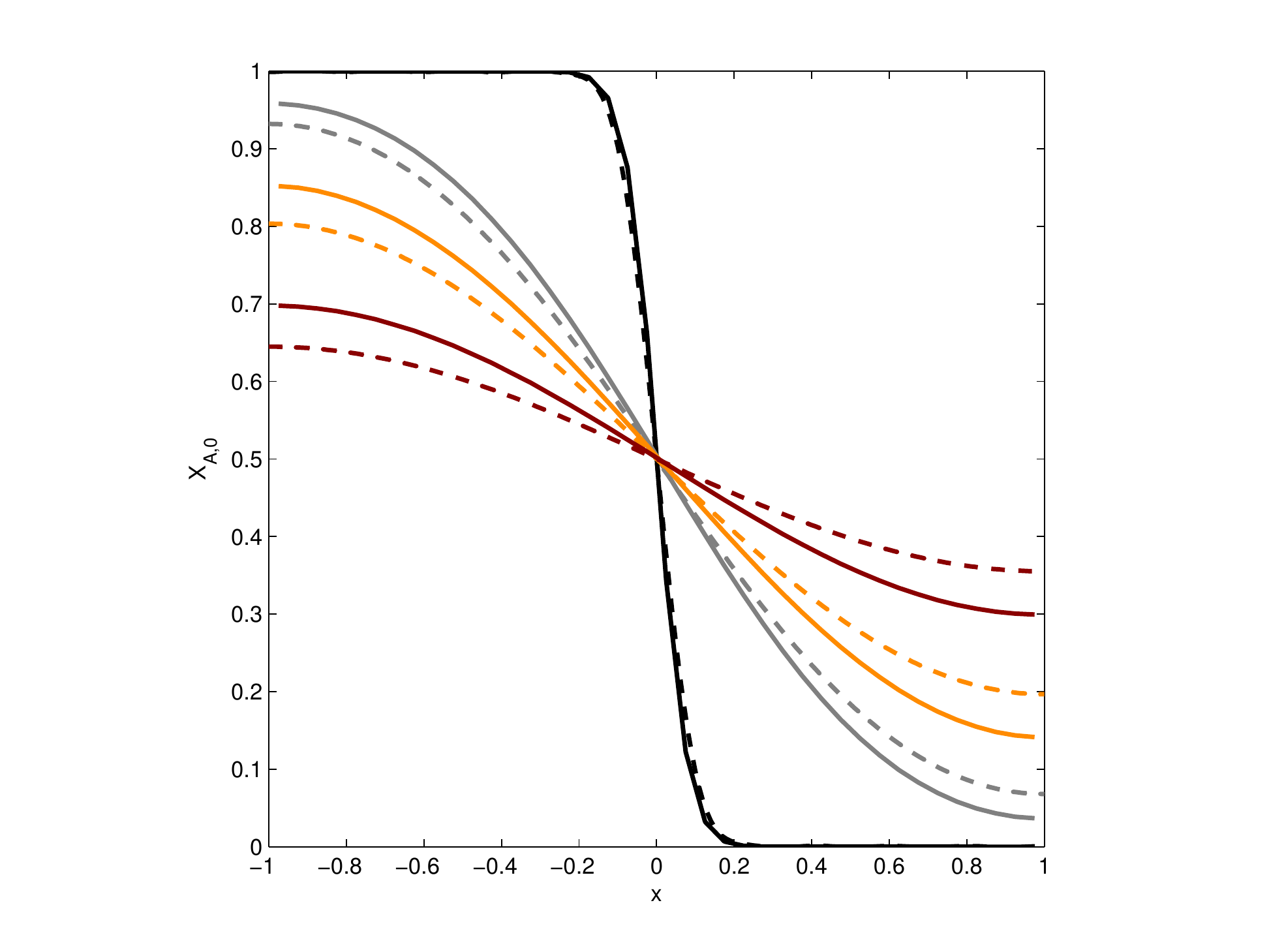}
\caption{$X_{A,0}$ for $\Gamma = 1.5$.}
\end{subfigure}%
\begin{subfigure}[b]{.5\textwidth}
\centering
\includegraphics[width=0.94\textwidth]{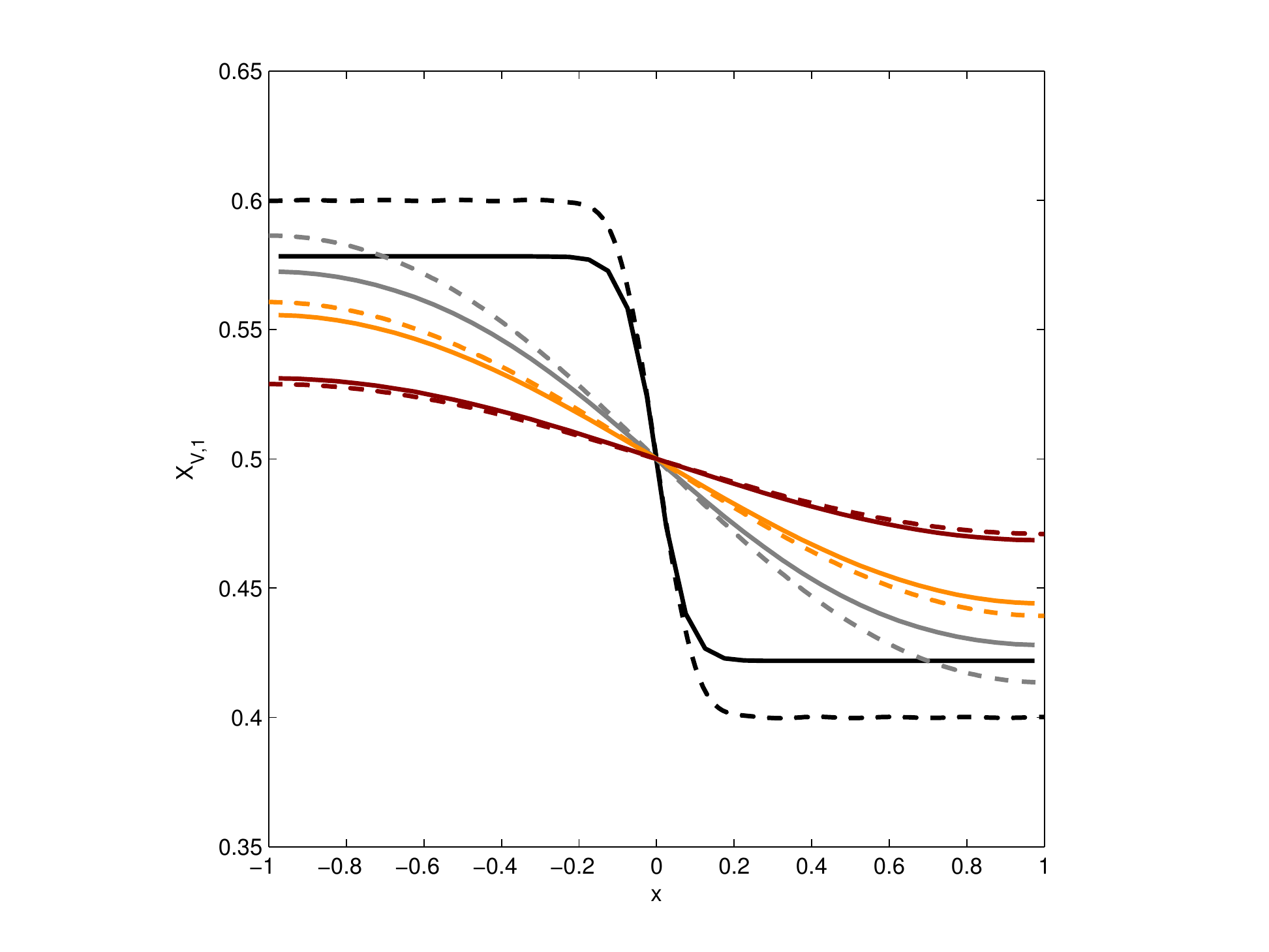}
\caption{$X_{V,1}$ for $\Gamma = 1.5$.}
\end{subfigure}%
}
\caption{$X_{A,0}$ and $X_{V,1}$ as given by the numerical solution to the full problem (solid) and by the analytical solution to the reduced problem $\Gamma \gg 1$, equations \eqref{new-sys-1-tau-2-u-analytical} and \eqref{new-sys-1-tau-2-v-analytical} (dashed), for different $\Gamma$ values. Different colours indicate different times: $t_1$ (black) $<$ $t_2$ (gray) $<$ $t_3$ (orange) $<$ $t_4$ (dark red).}
\label{figure-SlowTime-comparison}
\end{figure}

In Figure \ref{figure-SlowTime-comparison} we compare the numerical solution (solid) of $X_{A,0}$ (left) and $X_{V,1}$ (right) as described in Section \ref{sec-numerics} to the large $\Gamma$ approximate analytic solution (dashed) given by equations \eqref{new-sys-1-tau-2-u-analytical} and \eqref{new-sys-1-tau-2-v-analytical}, respectively. We take $10$ terms in the series to plot the solutions.
We choose 
$\Gamma = 10$ (Figure \ref{figure-SlowTime-comparison}(a), \ref{figure-SlowTime-comparison}(b)) and $\Gamma = 1.5$ (Figure \ref{figure-SlowTime-comparison}(c), \ref{figure-SlowTime-comparison}(d)).
As expected, for 
$\Gamma=10$, 
there is excellent agreement between the numerical and analytical solutions. The agreement deteriorates as $\Gamma$ decreases, however, even when $\Gamma=1.5$ the difference in $X_A$ on the interval $x \in [-1, 0]$ is only around 7\%. On the interval $[0,1]$ the difference is larger, this is a result of the simplification of the diffusion coefficients under large $\Gamma$ (equation \eqref{reduced-diffusion}) which  is valid provided $X_{A,0}$ is not close to zero. The error is most noticeable for small times and large $\Gamma$ near $x = 1$. However,  as time increases and so does $X_A$ the error also decreases. So, despite the fact that the simplification requires the assumptions that $\Gamma \gg 1$ and $X_{A,0}$ is not close to zero, the errors for 
$\Gamma = 1.5$ 
are reasonable
(Figure \ref{figure-SlowTime-comparison}(c), \ref{figure-SlowTime-comparison}(d)).
In all cases for sufficiently large times the solutions tend to equilibrium, that is, $X_{A,0} = X_{V,1} = 0.5$. To give an idea of the process time-scale we note that when $l = 10$ nm in the figures $t_4 = 366$ s, when $l = 10 \micrometer$, $t_4 = 3.66\times 10^8$ s.

\begin{figure}[!htb]
\makebox[\linewidth][c]{%
\begin{subfigure}[b]{.5\textwidth}
\centering
\includegraphics[width=0.94\textwidth]{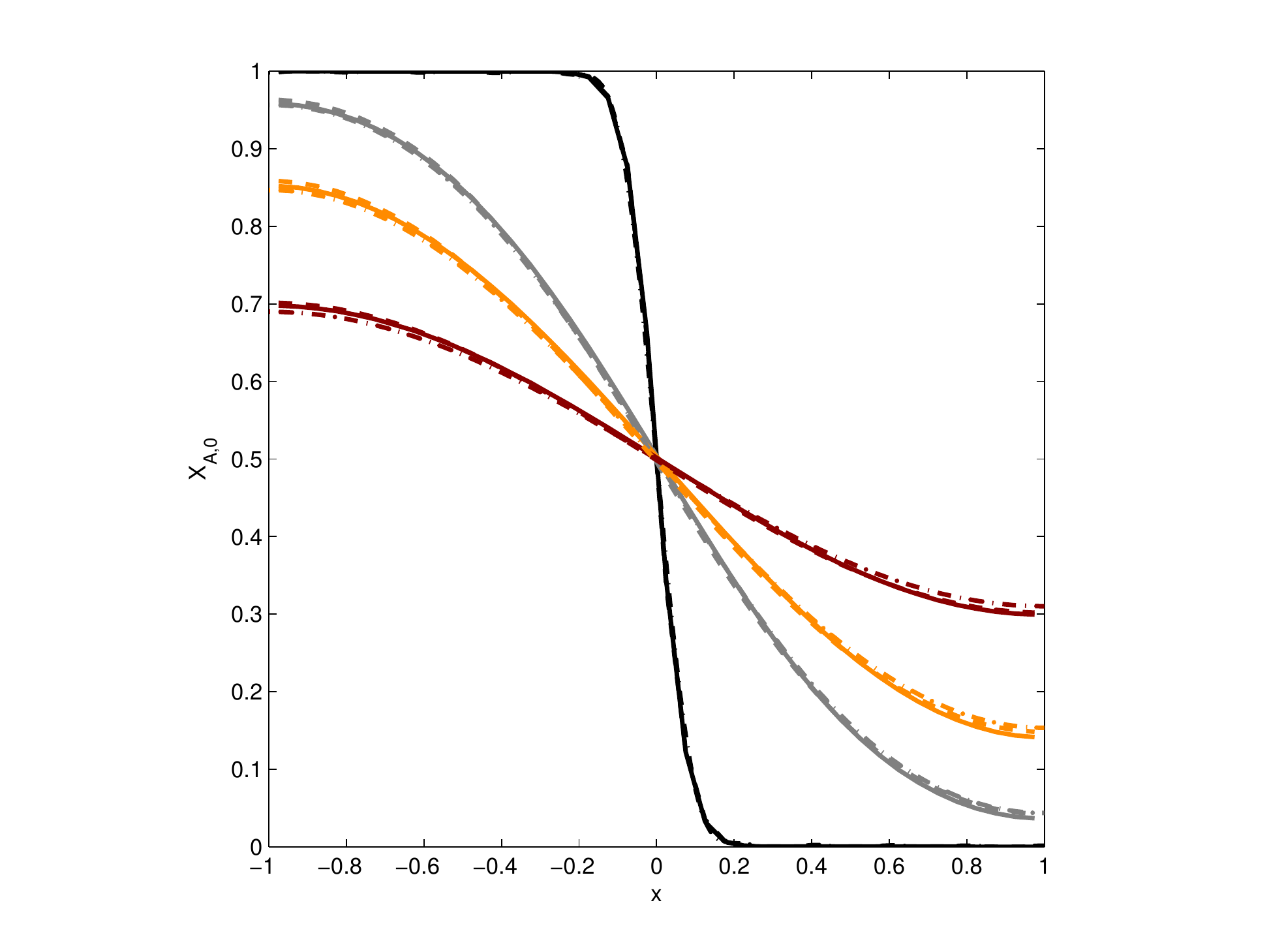}
\caption{$X_{A,0}$ for $\Gamma = 1.5$.}
\end{subfigure}%
\begin{subfigure}[b]{.5\textwidth}
\centering
\includegraphics[width=0.94\textwidth]{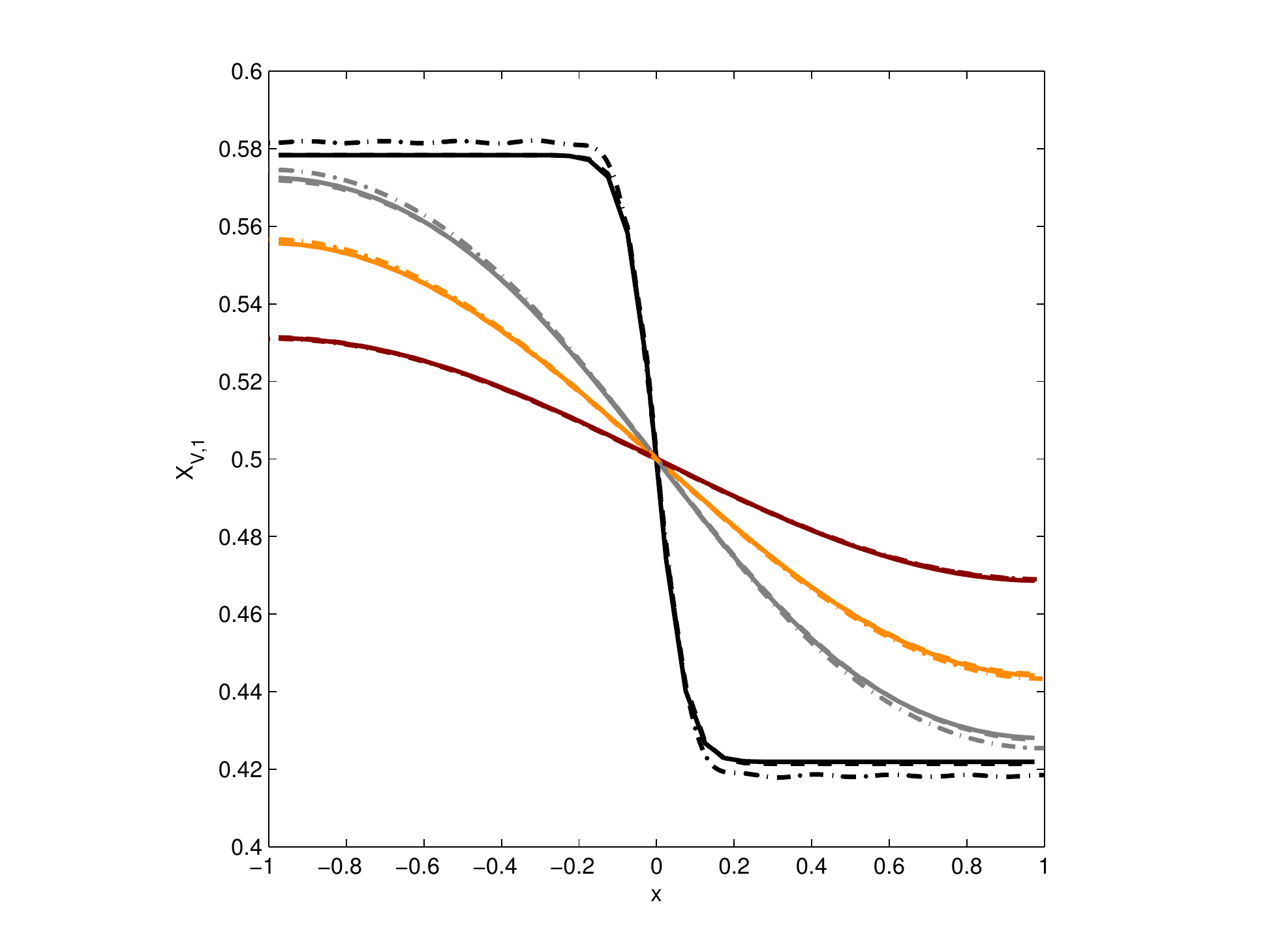}
\caption{$X_{V,1}$ for $\Gamma = 1.5$.}
\end{subfigure}%
}
\makebox[\linewidth][c]{%
\begin{subfigure}[b]{.5\textwidth}
\centering
\includegraphics[width=0.94\textwidth]{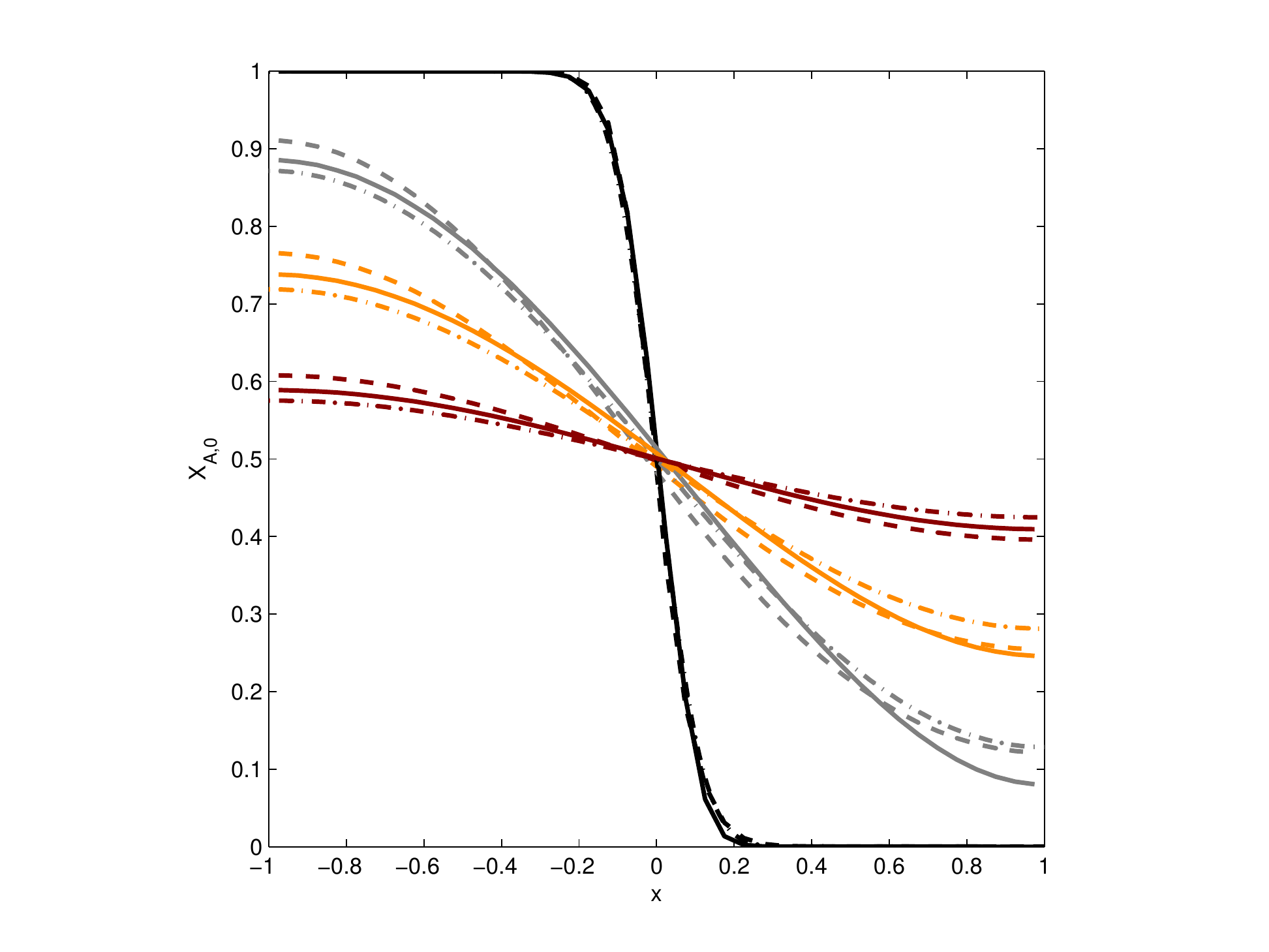}
\caption{$X_{A,0}$ for $\Gamma = 10$.}
\end{subfigure}%
\begin{subfigure}[b]{.5\textwidth}
\centering
\includegraphics[width=0.94\textwidth]{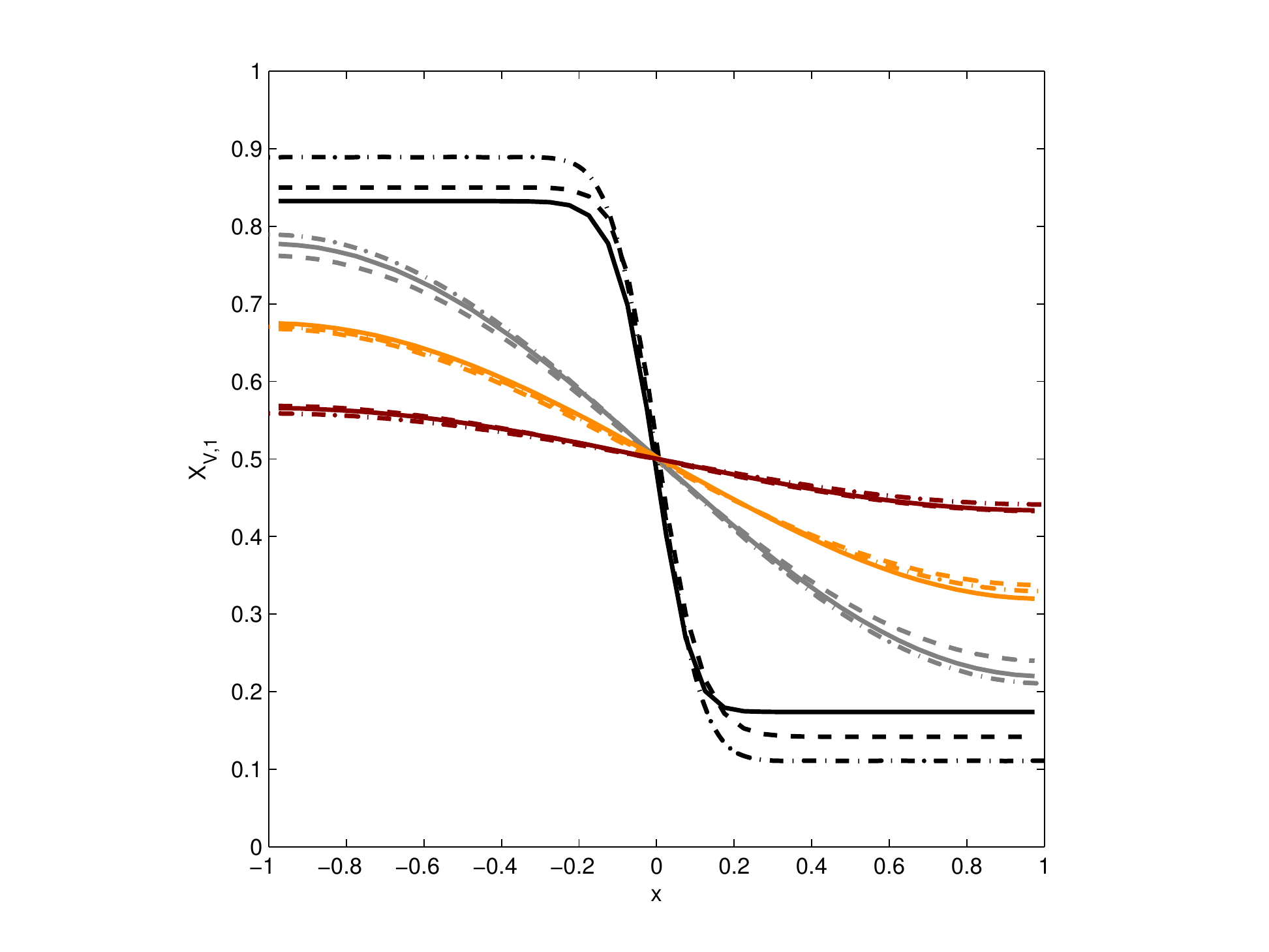}
\caption{$X_{V,1}$ for $\Gamma = 10$.}
\end{subfigure}%
}
\caption{Numerical solution of $X_{A,0}$ and $X_{V,1}$ to the full problem (solid), numerical solution to the reduced problem $\Gamma \sim 1$ (dashed), and analytical solution to the reduced problem $\Gamma \sim 1$ using $D_{AV}$ and $D_{VV}$ in equations \eqref{new-sys-1-tau-2-u-analytical-G1} and \eqref{new-sys-1-tau-2-v-analytical-G1} (dash-dotted). Plots for different $\Gamma$ values are presented. Different colours indicate different times: $t_1$ (black) $<$ $t_2$ (gray) $<$ $t_3$ (orange) $<$ $t_4$ (dark red).}
\label{figure-SlowTime-comparison-G1Reduction}
\end{figure}

In Figure \ref{figure-SlowTime-comparison-G1Reduction} we show a comparison of the full numerical solution (dashed), a numerical solution for the reduction where $\Gamma \sim 1$ (equation \eqref{reduced-diffusion-G1}) (dashed) and the analytical solution obtained using the approximation to $D_{AV}$ and $D_{VV}$ of equation \eqref{reduced-diffusion-G12} (dash-dotted) and taking $10$ terms in the series. For the case where $\Gamma = 1.5$,
Figure \ref{figure-SlowTime-comparison-G1Reduction}(a) and \ref{figure-SlowTime-comparison-G1Reduction}(b), the agreement between all three solutions is excellent although the analytical solution shows a slight error in the vacancy curves. It is interesting that the solution with an approximate diffusion coefficients is so accurate, since the error in $D_{AV}, D_{VV}$ may be around  $20$\%, so we must assume that these coefficients do not have a large effect on the solution.
For larger $\Gamma = 10$ (Figure \ref{figure-SlowTime-comparison-G1Reduction}(c) and \ref{figure-SlowTime-comparison-G1Reduction}(d)), as expected, the discrepancy increases.

\section{Conclusions}

One of the main aims of this paper was to derive governing equations for substitutional binary diffusion, which may then be used in different geometries since the continuity equations {\eqref{gov-eq-ficks-1}}, {\eqref{gov-eq-ficks-2}} are written in terms of the divergence and so different geometries, such as spherically symmetric can be directly obtained. Boundary and initial conditions depend on the specific problem, thus they need to be written when the problem is defined in those different geometries. Once derived we then reduced the equations to simulate substitutional diffusion in a one-dimensional bar. Our results indicated two distinct time-scales for the process: an initial fast time-scale where vacancies rapidly redistribute, followed by a slow redistribution to the constant steady-state.

The derived diffusion coefficients turned out to be quite complex, making the governing equations highly nonlinear. Useful reductions were only possible in a limited number of cases. For $\Gamma \gg 1$, that is, one species diffuses much faster than the other we were able to obtain an analytical solution, via separation of variables. The reduction was based on the volume fraction of fast diffuser not being close to zero. In general results were excellent, except for at small times, near $x=1$ where initially the volume fraction is zero. However, these errors decreased with time. For $\Gamma \sim 1$ analytical progress was made by slightly modifying two of the diffusion coefficients to give a system  that could again be solved using separation of variables. Despite the fact that the error in diffusion coefficients could be close to 20\% the errors arising from this modification were small.

Finally, we have developed a model which can be used readily to implement in other geometries (such as spherically symmetric) or with different boundary conditions, opening the doors to model the creation of hollow nanostructures.



\section*{Acknowledgements}

HR and TM acknowledge that the research leading to these results has received funding from ``la Caixa'' Foundation and has been partially funded by the CERCA Programme of the Generalitat de Catalunya.  TM acknowledges Ministerio de Ciencia e Innovación grant MTM2014-5621. BW acknowledges an NSERC Canada research grant.


\end{document}